\documentclass{article}
\usepackage{amsmath}
\usepackage{amssymb}

\begin{document}

\begin{center}
\textbf{Systems of several first-order quadratic recursions whose evolution
is easily ascertainable }

\bigskip

Francesco Calogero

Physics Department, University of Rome "La Sapienza", Rome, Italy

Istituto Nazionale di Fisica Nucleare, Sezione di Roma 1, Italy

Istituto Nazionale di Fisica Matematica, Gruppo Nazionale di Fisica
Matematica, Italy

francesco.calogero@uniroma1.it, francesco.calogero@roma1.infn.it

\bigskip
\end{center}

\textbf{Abstract}

The evolution, as functions of the "ticking time" $\ell =0,1,2,...$, of the
solutions of the system of $N$ \textit{quadratic} recursions%
\begin{eqnarray*}
x_{n}\left( \ell +1\right) =c_{n}+\sum_{m=1}^{N}\left[ C_{nm}x_{m}\left(
\ell \right) \right] +\sum_{m=1}^{N}\left\{ d_{nm}\left[ x_{m}\left( \ell
\right) \right] ^{2}\right\} && \\
+\sum_{m_{1}>m_{2}=1}^{N}\left[ D_{nm_{1}m_{2}}x_{m_{1}}\left( \ell \right)
x_{m_{2}}\left( \ell \right) \right] ~,~~~n=1,2,...,N~, &&
\end{eqnarray*}%
featuring $N+N^{2}+N^{2}+N\left( N-1\right) N/2=N\left( N+1\right) \left(
N+2\right) /2$ ($\ell $-independent) coefficients $c_{n}$, $C_{nm}$, $d_{nm}$
and $D_{nm_{1}m_{2}}$, may be \textit{easily ascertained}, if these
coefficients are given, in terms of $N+N^{2}=N\left( N+1\right) $ \textit{a
priori arbitrary} parameters $a_{n}$ and $b_{nm}$, by $N\left( N+1\right)
\left( N+2\right) /2$ \textit{explicit} formulas provided in this paper.
Here $N$ is an \textit{arbitrary} \textit{positive integer}.

\bigskip

\textbf{Introduction}

In this paper we identify a \textit{subclass} of the general system of 
\textit{autonomous first-order recursions} featuring \textit{quadratic}
right-hand sides,%
\begin{eqnarray}
x_{n}\left( \ell +1\right) =c_{n}+\sum_{m=1}^{N}\left[ C_{nm}x_{m}\left(
\ell \right) \right] +\sum_{m=1}^{N}\left\{ d_{nm}\left[ x_{m}\left( \ell
\right) \right] ^{2}\right\} &&  \notag \\
+\sum_{m_{1}>m_{2}=1}^{N}\left[ D_{nm_{1}m_{2}}x_{m_{1}}\left( \ell \right)
x_{m_{2}}\left( \ell \right) \right] ~,~~~n=1,2,...,N~, &&  \label{1}
\end{eqnarray}%
whose solution can be \textit{essentially achieved}.

\textbf{Notation}. In the above eq. (1), and hereafter, the independent
variable $\ell $ is a nonnegative integer, $\ell =0,1,2,3,...$ (say, a
"ticking time"), $N$ is an \textit{arbitrary }positive integer, $x_{n}\left(
\ell \right) $ are the $N$ dependent variables, $n$ is a positive integer
ranging from $1$ to $N,$ $n=1,2,...,N$, and the $N+N^{2}+N^{2}+N\left(
N-1\right) N/2=N\left( N+1\right) \left( N+2\right) /2$ coefficients $c_{n}$%
, $C_{nm}$, $d_{nm}$, $D_{nm_{1}m_{2}}$\ are $\ell $-independent. And please
also note that in this paper there are some minor notational changes with
respect to my previous $4$ papers (see Refs. [1,2,3,4]); and that it might
be considered just an extension of the paper [4]. Note moreover that in this
paper attention may be limited to the mathematics of \textit{real} numbers
(to the extent this is possible when dealing with \textit{nonlinear}
equations); indeed, it could even be restricted to the mathematics of 
\textit{real rational} numbers (since this is indeed possible when dealing
with recursions rather than differential equations). $\blacksquare $

This paper is a follow-up to my last $4$ papers put on \textbf{arXiv}
recently (see Refs. [1,2,3,4]), and it contradicts what is stated in the
last (Ref. [4]) of those $4$ papers, namely that that would have been my
last scientific paper; what motivated me to change my mind is the
realization that the simple findings mentioned just above and reported below
are sufficiently interesting to deserve to be shared; to be eventually
followed by \textit{examples }and possible \textit{applications}...

\bigskip

\textbf{Results}

The procedure to produce these findings was already mentioned in Ref. [4].
The starting point of our treatment are now just $N$ copies of the very
simple \textit{first-order nonlinear} recursion treated in Ref. [1]:%
\begin{equation}
y_{n}\left( \ell +1\right) =\left[ y_{n}\left( \ell \right) -1\right] ^{2}~.
\label{2}
\end{equation}

Next we introduce the following simple (\textit{linear}) change of variables
from the $N$ variables $y_{n}\left( \ell \right) $ to the $N$ dependent
variables $x_{n}\left( \ell \right) $: 
\begin{subequations}
\label{3}
\begin{equation}
x_{n}\left( \ell \right) =a_{n}+\sum_{m=1}^{N}\left[ b_{nm}y_{m}\left( \ell
\right) \right] ~,~~~n=1,2,...,N~.  \label{3a}
\end{equation}

Above and hereafter $a_{n}$ and $b_{nm}$ are $N+N^{2}=N\left( N+1\right) $ 
\textit{a priori} arbitrary parameters. It is notationally convenient to
also introduce the $N\times N$ matrix $\mathbf{B}$ of elements $b_{nm}$ and
the $N$-vectors $\mathbf{x}\left( \ell \right) $, $\mathbf{a}$, $\mathbf{y}%
\left( \ell \right) $, of components $x_{n}\left( \ell \right)
,~a_{n},~y_{n}\left( \ell \right) $, entailing that (3a) becomes the $N$%
-vector formula%
\begin{equation}
\mathbf{x}\left( \ell \right) =\mathbf{a+By}\left( \ell \right) ~,
\label{3b}
\end{equation}%
which is immediately inverted to read%
\begin{equation}
\mathbf{y}\left( \ell \right) =\mathbf{B}^{-1}\left[ \mathbf{x}\left( \ell
\right) -\mathbf{a}\right] ~.  \label{3c}
\end{equation}

Here and hereafter it is of course assumed that the matrix $\mathbf{B}$ is
invertible, namely that its $N^{2}$ \textit{a priori} arbitrary elements $%
b_{nm}$ satisfy the single constraint%
\begin{equation}
\det \left[ \mathbf{B}\right] \neq 0~;  \label{3d}
\end{equation}%
and we hereafter use the notation $\widetilde{\mathbf{B}}$ to denote the 
\textit{inverse} of the $N\times N$ matrix $\mathbf{B}$ and by $\widetilde{b}%
_{m_{1}m_{2}}$ its $N^{2}$ elements,%
\begin{equation}
\mathbf{B}^{-1}=\widetilde{\mathbf{B}}~,~~~\left( \widetilde{\mathbf{B}}%
\right) _{m_{1}m_{2}}=\widetilde{b}_{m_{1}m_{2}}~,  \label{3e}
\end{equation}%
and we denote by $\widetilde{\mathbf{c}}$ the $N$-vector $\mathbf{B}^{-1}%
\mathbf{a=}\widetilde{\mathbf{B}}\mathbf{a}$ and by $\widetilde{c}_{n}$ its $%
N$ components\textbf{,}%
\begin{equation}
\widetilde{\mathbf{c}}\mathbf{=B}^{-1}\mathbf{a=}\widetilde{\mathbf{B}}%
\mathbf{a~,~~~}\widetilde{c}_{n}=\sum_{m=1}^{N}\left( \widetilde{b}%
_{nm}a_{m}\right) ~.  \label{3f}
\end{equation}%
Hence eq. (3c) reads now, componentwise, as follows: 
\end{subequations}
\begin{subequations}
\label{4}
\begin{equation}
y_{n}\left( \ell \right) =\sum_{m=1}^{N}\left[ \widetilde{b}_{nm}x_{m}\left(
\ell \right) \right] -\widetilde{c}_{n}\ ,  \label{4a}
\end{equation}%
while eq. (3a) implies of course 
\begin{equation}
x_{n}\left( \ell +1\right) =a_{n}+\sum_{m=1}^{N}\left[ b_{nm}y_{m}\left(
\ell +1\right) \right] ~;  \label{4b}
\end{equation}%
while (2) implies via (4a), 
\end{subequations}
\begin{subequations}
\label{5}
\begin{eqnarray}
&&y_{n}\left( \ell +1\right) =\left\{ \sum_{m=1}^{N}\left[ \widetilde{b}%
_{nm}x_{m}\left( \ell \right) \right] -\widetilde{c}_{n}\right\} ^{2}  \notag
\\
&&-2\sum_{m=1}^{N}\left[ \widetilde{b}_{nm}x_{m}\left( \ell \right) \right]
+2\widetilde{c}_{n}+1~,  \label{5a}
\end{eqnarray}%
hence 
\begin{eqnarray}
&&y_{n}\left( \ell +1\right) =\sum_{m=1}^{N}\left[ \widetilde{b}%
_{nm}x_{m}\left( \ell \right) \right] ^{2}+2\sum_{m_{1}>m_{2}=1}^{N}%
\widetilde{b}_{nm_{1}}\widetilde{b}_{nm_{2}}x_{m_{1}}\left( \ell \right)
x_{m_{2}}\left( \ell \right)  \notag \\
&&-2\left( 1+\widetilde{c}_{n}\right) \sum_{m=1}^{N}\left[ \widetilde{b}%
_{nm}x_{m}\left( \ell \right) \right] +\left( \widetilde{c}_{n}+1\right)
^{2}~.  \label{5b}
\end{eqnarray}%
\bigskip

It is then easily seen, via the eqs. (2), (3), (4) and (5), that the $\ell $%
-evolution of the $N$ components of the $N$-vector $\mathbf{x}\left( \ell
\right) $ are indeed given by the system of recursions (1), with the
following definitions---in terms of the $N\left( N+1\right) $ \textit{a
priori} arbitrary parameters $a_{n}$ and $b_{nm}$---of the $N\left(
N+1\right) \left( N+2\right) /2$ coefficients $c_{n}$, $C_{nm}$, $d_{nm}$, $%
D_{nm_{1}m_{2}}$ featured by that system of recursions: 
\end{subequations}
\begin{subequations}
\label{6}
\begin{equation}
c_{n}=a_{n}+\sum_{m=1}^{N}\left[ b_{nm}\left( \widetilde{c}_{m}+1\right) ^{2}%
\right] ~,  \label{6a}
\end{equation}%
\begin{equation}
C_{nm}=-2\sum_{m_{1}=1}^{N}\left( b_{nm_{1}}\widetilde{b}_{m_{1}m}\left( 
\widetilde{c}_{m_{1}}+1\right) \right) ~,  \label{6b}
\end{equation}%
\begin{equation}
d_{nm}=\sum_{m_{1}=1}^{N}\left[ b_{nm_{1}}\left( \widetilde{b}%
_{m_{1}m}\right) ^{2}\right] ~,  \label{6c}
\end{equation}%
\begin{equation}
D_{nm_{1}m_{2}}=2\sum_{m=1}^{N}\sum_{m_{1}>m_{2}=1}^{N}\left( b_{nm}%
\widetilde{b}_{mm_{1}}\widetilde{b}_{mm_{2}}\right) ~.  \label{6d}
\end{equation}

Because we know a lot---see Ref. [1]---about the behavior of the solutions
of the simple recursion (2), we may easily infer a lot of detailed
informations about the behavior of the solutions of the subclass of the
system (1) on which we focus in this paper, namely that characterized by $%
N\left( N+1\right) \left( N+2\right) /2$ coefficients $c_{n}$, $C_{nm}$, $%
d_{nm}$, $D_{nm_{1}m_{2}}$ given in terms of the $N\left( N+1\right) $ 
\textit{a priori arbitrary} parameters $a_{n}$ and $b_{nm}$ by the formulas
(6). Here we outline some of the more obvious informations obtainable in
this manner; anybody interested in additional informations---perhaps being
motivated by an \textit{applicative} application of the system of recursions
(1)---may delve more deeply in the results reported in ref. [1].

The first immediate conclusion is that there shall be $2^{N}$ solutions of
the system of recursions (1) which are \textit{periodic} with period $2$, $%
x_{n}\left( \ell +2\right) =x_{n}\left( \ell \right) $; namely, such that $%
x_{n}\left( \ell \right) =x_{n}\left( 0\right) $ for \textit{even} $\ell $, $%
x_{n}\left( \ell \right) =x_{n}\left( 1\right) $ for \textit{odd} $\ell $, 
\end{subequations}
\begin{subequations}
\label{7}
\begin{eqnarray}
x_{n}\left( \ell \right) &=&x_{n}\left( 0\right) \text{ for }\ell =2,4,6,...,
\notag \\
x_{n}\left( \ell \right) &=&x_{n}\left( 1\right) \text{ for }\ell =1,3,5,...;
\label{7a}
\end{eqnarray}%
where of course (see (1)) 
\begin{eqnarray}
x_{n}\left( 1\right) =c_{n}+\sum_{m=1}^{N}\left[ C_{nm}x_{m}\left( 0\right) %
\right] +\sum_{m=1}^{N}\left\{ d_{nm}\left[ x_{m}\left( 0\right) \right]
^{2}\right\} &&  \notag \\
+\sum_{m_{1}>m_{2}=1}^{N}\left[ D_{nm_{1}m_{2}}x_{m_{1}}\left( 0\right)
x_{m_{2}}\left( 0\right) \right] ~,~~~n=1,2,...,N~. &&  \label{7b}
\end{eqnarray}%
They emerge from the $2^{N}~$sets of \textit{initial} data $x_{n}\left(
0\right) =\overline{x}_{n}^{\left( s\right) }$ with $s=1,2,...,2^{N}$ given
by the formulas 
\end{subequations}
\begin{equation}
x_{n}\left( 0\right) =\overline{x}_{n}^{\left( s\right)
}=a_{n}+\sum_{m=1}^{N}\left( b_{nm}\eta _{m}^{\left( s\right) }\right)
~,~~~n=1,2,3,,...,N~,  \label{8}
\end{equation}%
where$~$(above and hereafter) $\eta _{m}^{\left( s\right) }=0~$if $s$ is 
\textit{even} while $\eta _{m}^{\left( s\right) }=1~$if $s$ is \textit{odd. }%
This formula obviously provides generally $2^{N}$ \textit{different}
assignments for the number $\overline{x}_{n}^{\left( s\right) }$, to which
there shall correspond $2^{N}$ corresponding values for 
\begin{eqnarray}
x_{n}\left( 1\right) =\widetilde{x}_{n}^{\left( s\right)
}=c_{n}+\sum_{m=1}^{N}\left[ C_{nm}\overline{x}_{m}^{\left( s\right) }\left(
0\right) \right] +\sum_{m=1}^{N}\left\{ d_{nm}\left[ \overline{x}%
_{m}^{\left( s\right) }\left( 0\right) \right] ^{2}\right\} &&  \notag \\
+\sum_{m_{1}>m_{2}=1}^{N}\left[ D_{nm_{1}m_{2}}\overline{x}_{m_{1}}^{\left(
s\right) }\left( 0\right) \overline{x}_{m_{2}}^{\left( s\right) }\left(
0\right) \right] ~,~~~n=1,2,...,N~,~~~s=1,2,...,2^{N}~. &&  \label{9}
\end{eqnarray}

Moreover, for any set of \textit{other }initial data $x_{n}\left( 0\right) $
which are \textit{instead} all situated \textit{inside} the intervals of
oscillation of a periodic solution, namely such that, for all $n=1,2,...,N,$%
\begin{subequations}
\label{10}
\begin{eqnarray}
\overline{x}_{n}^{\left( s\right) } &<&x_{n}\left( 0\right) <\widetilde{x}%
_{n}^{\left( s+1\right) }~~~\text{if~~~}\overline{x}_{n}^{\left( s\right) }<%
\widetilde{x}_{n}^{\left( s+1\right) }~,  \notag \\
\widetilde{x}_{n}^{\left( s+1\right) } &<&x_{n}\left( 0\right) <\overline{x}%
_{n}^{\left( s\right) }~~~\text{if}~~~\widetilde{x}_{n}^{\left( s+1\right) }<%
\overline{x}_{n}^{\left( s\right) }~,  \label{10a}
\end{eqnarray}%
there shall hold the property of \textit{asymptotic isochrony} with period $%
2 $: 
\begin{equation}
x_{n}\left( \ell +2\right) -x_{n}\left( \ell \right) \rightarrow 0~~\ \text{%
as~~~}\ell \rightarrow \infty ~,  \label{10b}
\end{equation}%
with each component $x_{n}\left( \ell \right) $ of the solution jumping at
each step closer to one, and then to the other, of the $2$ borders of the
intervals (10a).

The interested reader may get additional informations about the evolution of
the solutions of the system (1) when its $N\left( N+1\right) \left(
N+2\right) /2$ coefficients $c_{n}$, $C_{nm}$, $d_{nm}$, $D_{nm_{1}m_{2}}$
are given in terms of the $N\left( N+1\right) $ \textit{a priori} arbitrary
parameters $a_{n}$ and $b_{nm}~$by~the $N\left( N+1\right) \left( N+2\right)
/2$ formulas (6), by utilizing some of the additional informations provided
in Ref. [1] on the behavior of the solutions of the simple single nonlinear
recursion (2).

Finally let us display a small Table displaying the $6$ values of the number 
$N\left( N+1\right) $ of freely assignable parameters $a_{n}$ and $b_{nm}$
and of the number $N\left( N+1\right) \left( N+2\right) /2$ of coefficients $%
c_{n}$, $C_{nm}$, $d_{nm}$ and $D_{nm_{1}m_{2}}$ of the system of recursions
(1), corresponding to the first $6$ positive integers $N$: 
\end{subequations}
\begin{eqnarray}
&&N:~\ \ \ \ \ \ \ \ \ \ \ \ \ \ \ \ \ \ \ \ \ \ \ \ \ ~1,~~~2,~~\
~~~3,~~~~~~4,~~~~~~5,~~~~~~~~6,...  \notag \\
&&N\left( N+1\right) :~~~~~~~~\ \ \ \ \ \ \ \
2,~~~~6,~~~~12,~~~~20,~~~~~30,~~~~~~42,...  \notag \\
&&N\left( N+1\right) \left( N+2\right)
/2:~~3,~~~12,~~~30,~~~~~60,~~~~~75,~~~~~~98,...~~~.  \label{11}
\end{eqnarray}

\bigskip

\textbf{Additional finding}

It is easy to see---by just drawing a graph of the $2$ sides of the \textit{%
algebraic} equation%
\begin{equation}
\overline{y}=\left( \overline{y}-1\right) ^{p}~,  \label{12}
\end{equation}%
whose solutions identify the equilibria $\overline{y}$ of the more general
class of recursions%
\begin{equation}
y_{n}\left( \ell +1\right) =\left[ y_{n}\left( \ell \right) -1\right] ^{p}~,
\label{13}
\end{equation}%
where $p$\ is now an \textit{arbitrary positive integer} larger than $2$, $%
p=3,4,5,...$---that for $p\ $\textit{any even positive integer}, $%
p=2,4,,6... $, there are only $2$ \textit{real} equilibrium solutions $%
\overline{y}$ of this algebraic equation (of degree $p$), one falling inside
the interval $0<\overline{y}<1$, and the other falling inside the interval $%
2<\overline{y}<3$ ; while for $p\ $\textit{any odd positive integer larger
than }$2$, $p=3,4,5,...$, there is only a \textit{single real} solution of
this algebraic equation, falling inside the interval $2<\overline{y}<3$.
Hence it may be easily seen that the behavior of \textit{all} real solutions
of the class of recursions (12) with $p=3,4,5,...$ is quite analogous to
that described in Ref. [1] for the case $p=2$; whenever the initial datum $%
y\left( 0\right) $ falls in the interval $0\leq y\left( 0\right) \leq 1$,
and as well when the initial datum falls \textit{outside} that interval
hence the solutions diverge as $\ell \rightarrow \infty $. And these
findings may of course be extended---as done above---to more general
recursions involving more than just $1$ dependent variable, via an analogous
change of variables to that discussed above (see (3a)); a task we leave for
the moment to whoever might be interested---maybe in view of its eventual 
\textit{applicative} relevance--to further explorations of these systems of
nonlinear recursions.

\bigskip

\textbf{References}

[1] F. Calogero, "Simple recursions displaying interesting evolutions",

arXiv2405.00370v1 [nlin.SI] 1 May 2024.

[2] F. Calogero, "Solvable nonlinear system of 2 recursions displaying
interesting evolutions",

arXiv:2407.18270v1 [nlin.SI] 20 Jul 2024.

[3] F. Calogero, "Interesting system of 3 first-order recursions",

arXiv:2409.05074v1 [nlin.SI] 8 Sep 2024.

[4] F. Calogero, "A simple approach to identify systems of nonlinear
recursions featuring solutions whose evolution is explicitly ascertainable
and which may be asymptotically isochronous as functions of the independent
variable (a ticking time)",

arXiv:2410.14448v1 [nlin.SI] 18 Oct 2024.

\end{document}